\documentclass[aps,prl,superscriptaddress,showkeys,showpacs,twocolumn,longbibliography,nofootinbib]{revtex4-1}

\usepackage[utf8]{inputenc}
\usepackage{amsfonts}
\usepackage{amsthm}
\usepackage{amssymb}
\usepackage{float}
\usepackage{braket}
\usepackage{graphicx}
\usepackage{url}
\usepackage[english]{babel}

\usepackage{tikz}
\usetikzlibrary{arrows.meta, shapes.geometric, positioning, chains, shapes, shadows, calc, decorations.pathmorphing, shapes.misc, math}

\usepackage{xcolor}

\usepackage[letterpaper,top=2cm,bottom=2cm,left=3cm,right=3cm,marginparwidth=1.75cm]{geometry}

\usepackage{amsmath}
\usepackage{graphicx}
\usepackage[colorlinks=true, allcolors=blue]{hyperref}

\begin{document}

\title{Fusion in a nanoshell: Harnessing plasmonic fields for nuclear reactions}

\author{\mbox{Dmitri E. Kharzeev}}
\email[]{dmitri.kharzeev@stonybrook.edu}

\affiliation{Center for Nuclear Theory, Department of Physics and Astronomy, Stony Brook University, Stony Brook, New York 11794-3800, USA}
\affiliation{Energy and Photon Sciences Directorate, Condensed Matter and Materials Sciences Division,
Brookhaven National Laboratory, Upton, New York 11973-5000, USA}

\author{Jacob Levitt}
\email[]{jakelevitt@cortexfusion.systems}
\affiliation{Cortex Fusion Systems, Inc., New York, NY 10128, USA}

\author{\mbox{Carlos Trallero-Herrero}}
\email[]{carlos.trallero@uconn.edu}

\affiliation{
Physics Department, University of Connecticut, Storrs, Connecticut 06269, USA}

\bibliographystyle{unsrt}

\begin{abstract}

The surface of metal nanoparticles can support plasmonic excitations. These excitations dramatically amplify the electric field of incident light (by several orders of magnitude), potentially ionizing the irradiated nanoparticles in a strong field regime.  Under specific conditions, a resonant enhancement of the electric field \textit{inside} a hollow nanoshell can be achieved with a laser pulse. We propose moderate-intensity laser irradiation of heavy water (${\rm D_2 O}$)-filled metal nanoshells to induce nuclear fusion via this enhancement. In this ``plasmonic confinement'' setup, deuteron nuclei are accelerated by the oscillating electric field within the nanoshell. We estimate that the characteristic momenta of the colliding deuterons reach approximately 10 MeV.  This corresponds to an effective kinetic energy equivalent to that of deuterons in a thermonuclear plasma at temperatures around 25 keV (approximately $10^8$K). A laser with a peak pulse intensity of roughly 1 atomic unit is sufficient to generate the strong electric fields required for this plasmonic acceleration. We estimate the expected fusion rate and discuss the feasibility of a fusion reactor based on this proposed scheme.


\end{abstract}

\maketitle

\noindent {\it Introduction.}  Achieving controlled nuclear fusion is one of the grand challenges of modern science. Under normal conditions, the Coulomb repulsion barrier prevents the nuclei from fusing. Two main ideas to achieve fusion  are vigorously pursued at present: magnetic confinement and inertial confinement, see \cite{ongena2016magnetic,betti2016inertial} for reviews. In both cases, the tunneling through the Coulomb barrier is enhanced by creating a hot plasma, and thus increasing the average kinetic energy of the colliding nuclei. 

Another, less known, approach to inducing nuclear fusion is Inertial Electrostatic Fusion (IEC) in which the nuclei are accelerated and confined by strong electrostatic fields, see \cite{miley2014inertial} for a review. The IEC method led to the first commercially available fusion devices, with rates of neutron production reaching $\sim 10^{10}\ {\rm n/s}$ \cite{Gradel}. While this rate is sufficient for practical and compact neutron sources, it is not enough for energy generation. The factors limiting the fusion rate include the limit on the achievable electric field, its inhomogeneities,
the deuteron loss to electrodes, and instabilities.

In this letter, we propose to overcome some of these limitations by leveraging the strong electric fields inside metallic nanoshells induced by moderately intense pulsed lasers. The electric field of the laser light in this case is known to be dramatically enhanced by the collective plasmon excitations excited on the surface of metallic nanoparticles \cite{powellStrongFieldControlPlasmonic2022}. If the interior of a metallic nanoshell is filled with deuterium nuclei (for example, bound inside the heavy water molecule ${\rm D_2 O}$), these nuclei will be subjected to a strong electric field that oscillates in the direction perpendicular to the surface of the nanoshell. The deuteron nuclei will thus be accelerated towards each other and collide in the interior of the nanoshell, see Fig. 1.
\vskip0.3cm

\noindent {\it Electrodynamics of metal nanoshells.} The idea of using electric fields inside a metal nanoshell to accelerate the deuterons is in apparent contradiction with two facts known in classical electromagnetism: i) the electric field inside a uniformly charged sphere vanishes; ii) Faraday shield (cage) effect that protects the interior of a conducting shell from external electric fields. However, neither of these contradictions prevent the existence of strong electric fields inside the metal nanoshells in our setup:

i) the plasmon excitation creates an oscillating and highly inhomogeneous distribution of electric charge on the surface of the nanoshell. In this case, the electric field inside the nanoshell does not vanish, as can be seen from the Gauss theorem;

ii) the thickness of the nanoshell can be smaller than the ``skin depth" that determines the efficacy of the Faraday shielding \cite{Jackson3rded}. 
\vskip0.3cm

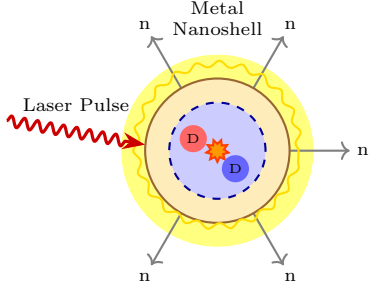
\begin{figure}
    \centering

             \begin{tikzpicture}[
        scale=0.8, node distance=0.5cm and 1cm,
        laser/.style={draw=red!70!black, very thick, -{Stealth[length=3mm]}},
        shell_outer/.style={thick, draw=brown!80!black, fill=yellow!60!orange!30, opacity=0.6},
        shell_inner/.style={thick, draw=blue!60!black, fill=blue!20, opacity=0.4, dashed}, 
        particle_p/.style={circle, fill=red!60, inner sep=1.5pt, font=\tiny}, 
        particle_O/.style={circle, fill=blue!60, inner sep=1.5pt, font=\tiny}, 
        emission/.style={draw=gray, thick, ->}, 
        glow/.style={inner color=yellow!50, outer color=yellow!50!white, opacity=0.3},
        fusion_flash/.style={star, star points=9, star point ratio=0.6, fill=orange!80!yellow, inner sep=3pt, draw=orange!50!red, thick}, 
        plasmon_deco/.style={decorate, decoration={snake, amplitude=0.4mm, segment length=3mm}, yellow!80!orange, thick} 
    ]
    \shade[glow] (0,0) circle (1.6cm);


\foreach \angle/\label in {0/n, 60/n, 120/n, 240/n, 300/n} {
        \pgfmathtruncatemacro{\angleint}{\angle} 
        \ifnum\angleint=180
            \def\myanchoropt{anchor=east} 
            \def\mypos{1.15} 
        \else
            \def\myanchoropt{} 
            \def\mypos{1.1} 
        \fi
        \draw[emission] (0,0) -- (\angle:2.2cm) node[pos=\mypos, font=\scriptsize, black, \myanchoropt] {\label};
    }

    \draw[shell_outer] (0,0) circle (1.2cm);
    \draw[shell_inner] (0,0) circle (0.8cm); 
    \node[font=\scriptsize, align=center] at (0,2.2cm) {Metal\\Nanoshell}; 

    \draw[plasmon_deco] (0,0) circle (1.35cm); 

    \node[fusion_flash] at (0,0) {}; 

    \coordinate (LaserStart) at (-3.5, 0.5);
    \coordinate (LaserEnd) at (-1.2, 0.1);
    \coordinate (SquiggleEnd) at ($(LaserStart)!0.9!(LaserEnd)$);

    \draw[red!80!black,
          very thick,
          decorate,
          decoration={snake,
                      amplitude=2pt,    
                      segment length=5pt
                     }
         ]
        (LaserStart) -- (SquiggleEnd); 

    \draw[red!80!black, very thick, -{Stealth[length=3mm]}] 
        (SquiggleEnd) -- (LaserEnd); 

    \node[font=\scriptsize, above, color=black, inner sep=1pt] at ($(LaserStart)!0.5!(LaserEnd) + (0,0.3)$) {Laser Pulse};
    \node[particle_p] (p1) at (-0.4, 0.2) {D}; 
    \node[particle_O] (O16) at (0.3, -0.3) {D}; 


\end{tikzpicture}
    \caption{A schematic representation of the deuteron fusion inside a gold nanoshell induced by the dipole electric field.}
    \label{fig:enter-label}
\end{figure}

A reasonable starting point for estimating the electric field inside a metal nanoshell is Mie's theory describing the scattering of a plane wave of light by a sphere. The existing code for solving Mie equations \cite{LADUTENKO2017225} assumes that the skin depth of the current is smaller than the thickness of the metal shell. In this case, the Faraday shielding applies, and the electric field inside the nanoshell indeed vanishes \cite{LADUTENKO2017225}.

However, in the thin-shell regime, where the shell thickness is comparable to or less than the skin depth, Faraday shielding breaks down. This allows the surface plasmon resonance to couple to all electrons in the shell, resulting in a strong, oscillating electric field inside the nanoshell driven by the dipole plasmon resonance. We should point out that experimental evidence of Faraday shielding breaking down already exists for setups very similar to the one proposed here \cite{powellStrongFieldControlPlasmonic2022}.


We consider a hollow metallic spherical nanoshell filled with a heavy water ${\rm D_2 O}$. Let us denote the inner radius of the sphere $R_1$, the outer radius $R_2$, and assume that the dielectric permittivity inside the sphere is $\varepsilon_{\text{in}}$. The permittivity outside the sphere is $\varepsilon_{\text{out}}$, whereas in the metallic shell, the permittivity $\varepsilon_m(\omega)$ is frequency-dependent. All of these permittivities are complex.
\vskip0.3cm

We aim to derive the electric field $\mathbf{E}_{\text{in}} = - {\bf \nabla} \Phi_{\text{in}}$ inside the hollow sphere. In the electrostatic approximation, this can be done by writing the potentials $\Phi$ inside the sphere, in the shell, and outside the sphere that solve the corresponding Laplace equations, and matching the corresponding expressions for the potentials and displacement fields $\mathbf{D} = \varepsilon  \mathbf{E}$ at the boundaries $r=R_1$ and $r=R_2$. This problem was solved in  \cite{Neeves:89}, and we have reproduced their results. If the electric field of the laser ${\mathbf{E}_0}$ is linearly polarized along the $z$ axis, the electric field inside the sphere is given by 
\begin{equation}\label{elin}
    {\mathbf{E}} = \frac{9\ \varepsilon_m \varepsilon_{\text{out}}}{\varepsilon_m \varepsilon_a + 2 \varepsilon_{\text{out}} \varepsilon_b}\ E_0\ \left[ \cos \theta \  {\hat{e_r}} - \sin \theta \  {\hat{e_\theta} } \right], 
\end{equation}
where 
\begin{equation*}
    \varepsilon_a = \varepsilon_{\text{in}} (3 - 2 P) + 2 \varepsilon_m P,
\end{equation*}
\begin{equation*}
    \varepsilon_b = \varepsilon_{\text{in}} P +  \varepsilon_m (3 - P),
\end{equation*}
and 
\begin{equation*}
    P = 1 - \left(\frac{R_1}{R_2}\right)^3 .
\end{equation*}

The surface plasmon resonance condition is satisfied at a frequency $\omega_p$ when the real part of the denominator in (\ref{elin}) vanishes, 
\begin{equation}
\Re \left[ \varepsilon_m \varepsilon_a + 2 \varepsilon_{\text{out}} \varepsilon_b \right] = 0;
\end{equation}
the magnitude of the enhancement will then be determined by the corresponding imaginary part, $\Im [ \varepsilon_m \varepsilon_a + 2 \varepsilon_{\text{out}} \varepsilon_b  ]$.

At this frequency, we get a  strong plasmonic enhancement:
\begin{equation}
\mathbf{E}_{\text{in}} \gg \mathbf{E}_0 \quad \text{(near resonance)}.
\end{equation}

While direct measurement of the internal field is difficult, experimental observations have confirmed external field enhancements exceeding a factor of a thousand near gold nanoshells \cite{powellStrongFieldControlPlasmonic2022}. Theoretical considerations for the thin-shell regime, particularly the breakdown of Faraday shielding and the resulting oscillating electric dipole moment, strongly suggest a similar enhancement of the internal field.


\vskip0.3cm
\noindent {\it Plasmonic confinement.} Due to the large mass of deuterons, their motion is adequately described by classical equations of motion. The momentum $P$ gained by the deuteron is thus given by the Newton's law
\begin{equation}\label{eom}
    \frac{dP}{dt} = e E(t),
\end{equation}
where $E(t)$ is the time-dependent oscillating  of the total field, including the plasmonic, residual charges, and the external field, and $e$ is the charge of the deuteron (equal to the elementary charge).  

As a consequence of (\ref{eom}), the maximal momentum gained by the nuclei is limited by the period $\tau$ of the plasmon excitation, $P_0 \sim eE_0 \tau$, where $E_0$ is the maximal amplitude of the electric field. Since this period is quite short, on the order of a few femtoseconds, it limits the momentum achievable by the nuclei even in very strong plasmon fields that can be in excess of $E_0 \sim 10^{11}\ {\rm V/cm}$. 

For example, assuming $\tau=3$ fs and $E_0 = 10^{11}\ {\rm V/cm}$ (corresponding to the plasmon field inside the nanoshell induced by a $\simeq 1\ \mu$m wavelength laser field with a peak electric field of about $10^9\ {\rm V/cm}$), we get from (\ref{eom}) the momentum of 
\begin{equation}\label{momest}
    P_0 \simeq eE_0 \tau \simeq 10\ {\rm MeV}
\end{equation}
corresponding to the deuteron kinetic energy of $E_K\simeq P_0^2/2M_D \simeq 25$ KeV, where $M_D$ is the deuteron mass. 

To achieve such momenta and kinetic energies in a thermalized plasma, one would need to heat up the plasma to temperatures of $T \simeq 25$ KeV. For comparison, a typical plasma temperature in a plasma confined inside a Tokamak is $T \sim 10-30$ KeV. Indeed, according to the nanoplasma model \cite{ditmireInteractionIntenseLaser1996,fennelReview2010}, the condition to achieve temperatures of $T \gg 1\ {\rm keV}$ is satisfied for nano-clusters of size larger than the Debye length $\lambda_D > 0.5$ nm. Therefore achieving Tokamak-like temperatures inside the nanoshells of radius $R_2> 12.5$ nm is very likely. 
\vskip0.3cm

The ionization of the nanoshells by the strong plasmonic field, a process observed to remove thousands of electrons \cite{powellStrongFieldControlPlasmonic2022,ErfanStrongfieldIonizationPlasmonic2022,ErfanEnhancedCutoffEnergies2023,davino2025extreme}, has not yet been incorporated into our analysis. This ionization leads to a dampening of the plasmon oscillation. 
Further investigation is needed to fully understand this effect.

The electric field accelerates electrons to the same momentum (\ref{momest}), resulting in ultra-relativistic electrons with a very high effective temperature $T \simeq P_0/3 \simeq 3$ MeV.  The subsequent thermalization process within the nanoshell involves electron-deuteron scattering, which will drive the overall plasma temperature above the $T \simeq 60$ keV value estimated previously.  Furthermore, oxygen ions, due to their higher electric charge $Z=8$, will reach an even higher effective temperature, $T_O \simeq (8 P_0)^2/2M_O \simeq 200$ keV. As a result, the thermalized plasma inside the nanoshell is highly likely to possess an effective temperature exceeding $T \simeq 25$ keV.



\vskip0.3cm

\noindent {\it Equation of motion for the deuteron acceleration.} Above we have estimated the maximal momentum of the deuterons accelerated by the plasmon field. Now we have to take into account that this acceleration has to happen within the nanoshell, which is likely to reduce the estimate for the momentum given above. To correlate the momentum of the deuteron with its position inside the sphere, let us solve the equation of motion (\ref{eom}) with the initial condition $p(0)=0$, $x(0)=0$, i.e. we consider a deuteron which initially was at rest near the surface of the nanoshell. With these initial conditions, we find the trajectory (see e.g. \cite{Kharzeev:2005iz})
\begin{equation}\label{vel}
    v(t) = \frac{at}{\sqrt{1+a^2 t^2}}, 
\end{equation}
\begin{equation}\label{coord}
    x(t) = a^{-1}\ \left(\sqrt{1+a^2 t^2} - 1\right),
\end{equation}
where the acceleration $a$ is given by
\begin{equation}\label{acc}
    a = \frac{d}{dt} \frac{v}{\sqrt{1-v^2}} = \frac{e E_0}{M_D} .
\end{equation}
When $at\ll 1$, (\ref{vel}) and (\ref{coord}) reduce to the usual non-relativistic formulae  $v(t) \simeq at$, $x(t) \simeq at^2/2$.

Assuming as before the plasmon-enhanced electric field inside the nanoshell of  $E_0 = 10^{11}$ V/cm, 
The deuteron acceleration according to (\ref{acc}) is 
$a \simeq 10^{11}\ {\rm eV\ cm^{-1}} /(1.87\ 10^9\ {\rm eV})\simeq 50\ {\rm cm^{-1}}$.
During the period of the plasmon oscillation that we assume be equal to $\tau = 3$ fs, the deuteron will accelerate to the momentum 
(\ref{momest}). Its coordinate at that time according to (\ref{coord}) will be $x(\tau) \simeq 6$ nm -- well within a typical radius of the nanoshell, $R \sim {\it O}(100)\ {\rm nm}$. 

This confirms our previous statement that an effective temperature of the deuteron plasma inside the nanoshell should be very high,  $T \simeq 60$ KeV. The kinetics of thermalization during the interactions with even ``hotter" electrons and oxygen ions is likely to drive it even higher.  The corresponding temperature is thus likely higher than $T\simeq 10 - 30$ KeV typical for a Tokamak plasma.
\vskip0.3cm
\noindent {\it Fusion probability}.
The main factor limiting the fusion rate in our setup is the fact that even though the momenta of deuterons inside the nanoshell are high, and correspond to a high effective temperature, the density of deuterons is limited by the number of ${\rm D_2 O}$ molecules initially placed inside the nanoshell, 
which is relatively low compared to a conventional thermonuclear plasma.

To get a rough estimate of the probability of deuteron fusion, we have to consider the density of deuterons inside the nanoshell 
\begin{equation}
   \rho_D = 2\times 3.3\ 10^{28}\ {\rm m}^{-3} = 66\ {\rm nm}^{-3}.  
\end{equation}

The fusion rate is given by
\begin{equation}\label{rate}
    R_f \simeq \rho_D\ v\ \sigma^f_{DD},
\end{equation}
where $v \sim P_0/M_D$ is the characteristic relative deuteron velocity. This yields an estimate of 
\begin{equation}\label{rate1}
    R_f \sim 10^7\ {\rm s}^{-1}
\end{equation}
per nanoshell. 
\vskip0.3cm

The traditional estimate (\ref{rate}) may be questioned, as the deuteron mean free path w.r.t. fusion at the relevant low density exceeds the size of the nanoshell,  $\lambda_D^f = (\rho_D\  \sigma^f_{DD})^{-1} \sim {\rm cm}$. 
However, the deuterons will undergo multiple interactions with hot electrons inside the nanoshell. Indeed, we can estimate the mean free path of deuterons $\lambda_D$ as
\begin{equation}
\lambda_D \simeq \left(\frac{T}{e^2\ n_e}  \right)^{1/2}, 
\end{equation}
where $T \simeq (e E_0 \tau)/3 \simeq 3\ {\rm MeV}$ is the effective temperature of relativistic electron gas, and $n_e = 10 \times 3.3\ 10^{28}\ {\rm m}^{-3}$ is the electron density, which in our setup is much smaller than the equilibrium density at temperature $T$. Numerically, we find the mean free path of the deuteron of
$$
\lambda_D \simeq 10\ {\rm nm},
$$
which is smaller than the typical radius of the nanoshell.

However, a significant fraction of the accelerated deuterons will escape from the nanoshell. If the nanoshells are suspended in heavy water (${\rm D_2 O}$), these escaping deuterons can still contribute to fusion. The heavy water outside the nanoshells will be partially ionized by the plasmonic fields, so the electron distribution will have both ``cold" and ``hot" components. The relative weights of these components will depend on the volume fraction of nanoshells.  A detailed study of the relevant kinetics is required for  accurate estimates.

\vskip0.3cm

\noindent {\it Released energy}. 
To estimate the power produced by the proposed ``nano-Sun" fusion reactor we have to account for the fact that the laser light is pulsed, and that a limited number of nanoshells can be ignited per second. We envision pulsed laser irradiation of a heavy water-based nanoshell colloid with a high volume fraction of nanoshells.


Each $DD$ fusion releases $3$ MeV, or $2\ 10^{-13} J$ of energy, so the fusion rate (\ref{rate1}) corresponds to the power of 
\begin{equation*}
    P_{sph} \sim 1\ {\mu} W
\end{equation*}
produced by a single nanoshell.

If $N_{s-p}$ nanoshells are ignited per laser pulse, and the pulse repetition frequency is $f$, then the produced power is
\begin{equation}
    P = P_{sph}\ N_{s-p}\ f .
\end{equation}
To get a potentially practical fusion reactor with a laser repetition frequency of $f=1$ MHz, we thus need to ignite a million nanoshells per pulse -- this will produce the power output $P$ of 
\begin{equation}
    P_{fusion} \sim 1\ {\rm MW} .
\end{equation}
 The corresponding neutron flux $F$ will be 
 \begin{equation}
     F \sim 10^7 s^{-1} \cdot 10^6 \cdot 10^6\ {\rm n}\sim 10^{19} \ {\rm n/s},
 \end{equation}
exceeding the flux available from current fusion devices by about nine orders of magnitude. 
\vskip0.3cm

To evaluate the $Q$ factor of a possible nano-Sun reactor, we need to account for the efficiency $\kappa$ of conversion of the $\gamma$ quanta and neutron energy into electric power, which is about $\kappa \simeq 30 \%$, and the power consumed by the laser system, which is $P_{laser} \simeq 3$ KW. This implies a $Q$ factor on the order of 
\begin{equation}
   Q \equiv \frac{P_{fusion}\ \kappa}{P_{laser}}\simeq  100 
\end{equation} is theoretically achievable for the proposed nano-Sun reactor. 
\vskip0.3cm 
Although many practical challenges exist, our estimates suggest that achieving this magnitude of power production is likely within reach. Our optimism is grounded in current advances in commercial-grade Yb-based lasers, which are capable of delivering intense pico- and femtosecond pulses at repetition frequencies of hundreds of kHz \cite{beetar2020,watsonHighpowerFemtosecond2025}, as well as in the fact that nanoparticles maintain a highly energetic response throughout this entire temporal range \cite{davino2025extreme}.
\vskip0.3cm

\textit{Intellectual Property Notice}: The systems, methods, and underlying principles described in this paper are the intellectual property of Cortex Fusion Systems, Inc., and are protected by various intellectual property rights, including but not limited to patents and pending patent applications.
\vskip0.3cm

The work of D.K. was supported in part by the U.S. Department of Energy, Office of Science,
Office of Nuclear Physics, Grants No. DE-FG88ER41450 and DE-SC0012704. The work of C. T-H was partially supported by the Air  Force  Office  of  Scientific  Research, Grant No FA9550-21-1-0387.

\bibliography{main1}

\begin{thebibliography}{10}

\bibitem{ongena2016magnetic}
J~Ongena, R~Koch, R~Wolf, and H~Zohm.
\newblock Magnetic-confinement fusion.
\newblock {\em Nature Physics}, 12(5):398--410, 2016.

\bibitem{betti2016inertial}
R~Betti and OA~Hurricane.
\newblock Inertial-confinement fusion with lasers.
\newblock {\em Nature Physics}, 12(5):435--448, 2016.

\bibitem{miley2014inertial}
George~H Miley and S~Krupakar Murali.
\newblock Inertial electrostatic confinement (iec) fusion.
\newblock {\em Fundamentals and Applications}, pages 75--80, 2014.

\bibitem{Gradel}
NSD-Gradel~Fusion company:.
\newblock \ http://www.nsd-fusion.com.

\bibitem{powellStrongFieldControlPlasmonic2022}
Jeffrey~A. Powell, Jianxiong Li, Adam Summers, Seyyed~Javad Robatjazi, Michael
  Davino, Philipp Rupp, Erfan Saydanzad, Christopher~M. Sorensen, Daniel
  Rolles, Matthias~F. Kling, Carlos Trallero, Uwe Thumm, and Artem Rudenko.
\newblock Strong-{{Field Control}} of {{Plasmonic Properties}} in {{Core-Shell
  Nanoparticles}}.
\newblock {\em ACS Photonics}, 9(11):3515--3521, November 2022.

\bibitem{Jackson3rded}
John~David Jackson.
\newblock {\em Classical {{Electrodynamics}}, 3rd {{Edition}} {\textbar}
  {{Wiley}}}.
\newblock Wiley, New York, NY, 3rd edition edition, 1998.

\bibitem{LADUTENKO2017225}
Konstantin Ladutenko, Umapada Pal, Antonio Rivera, and Ovidio Peña-Rodríguez.
\newblock Mie calculation of electromagnetic near-field for a multilayered
  sphere.
\newblock {\em Computer Physics Communications}, 214:225--230, 2017.

\bibitem{Neeves:89}
A.~E. Neeves and M.~H. Birnboim.
\newblock Composite structures for the enhancement of nonlinear-optical
  susceptibility.
\newblock {\em J. Opt. Soc. Am. B}, 6(4):787--796, Apr 1989.

\bibitem{ditmireInteractionIntenseLaser1996}
T.~Ditmire, T.~Donnelly, A.~M. Rubenchik, R.~W. Falcone, and M.~D. Perry.
\newblock Interaction of intense laser pulses with atomic clusters.
\newblock {\em Physical Review A}, 53(5):3379--3402, May 1996.

\bibitem{fennelReview2010}
{\relax Th}.~Fennel, K.-H. {Meiwes-Broer}, J.~Tiggesb{\"a}umker, P.-G.
  Reinhard, P.~M. Dinh, and E.~Suraud.
\newblock Laser-driven nonlinear cluster dynamics.
\newblock {\em Reviews of Modern Physics}, 82(2):1793--1842, June 2010.

\bibitem{ErfanStrongfieldIonizationPlasmonic2022}
E.~Saydanzad, J.~Li, and U.~Thumm.
\newblock Strong-field ionization of plasmonic nanoparticles.
\newblock {\em Physical Review A}, 106(3):033103, September 2022.

\bibitem{ErfanEnhancedCutoffEnergies2023}
Erfan Saydanzad, Jeffrey Powell, Adam Summers, Seyyed~Javad Robatjazi, Carlos
  {Trallero-Herrero}, Matthias~F. Kling, Artem Rudenko, and Uwe Thumm.
\newblock Enhanced cutoff energies for direct and rescattered strong-field
  photoelectron emission of plasmonic nanoparticles.
\newblock {\em Nanophotonics}, 12(10):1931--1942, May 2023.

\bibitem{davino2025extreme}
Michael Davino, Tobias Saule, Nora~G Helming, and Carlos~A Trallero-Herrero.
\newblock Extreme pulse duration scaling of strong field ionization of
  nanoparticles.
\newblock {\em ACS Photonics}, 2025.

\bibitem{Kharzeev:2005iz}
Dmitri Kharzeev and Kirill Tuchin.
\newblock {From color glass condensate to quark gluon plasma through the event
  horizon}.
\newblock {\em Nucl. Phys. A}, 753:316--334, 2005.

\bibitem{beetar2020}
John~E. Beetar, M.~Nrisimhamurty, Tran-Chau Truong, Garima~C. Nagar, Yangyang
  Liu, Jonathan Nesper, Omar Suarez, Federico Rivas, Yi~Wu, Bonggu Shim, and
  Michael Chini.
\newblock Multioctave supercontinuum generation and frequency conversion based
  on rotational nonlinearity.
\newblock {\em Science Advances}, 6(34):eabb5375, August 2020.

\bibitem{watsonHighpowerFemtosecond2025}
Kevin Watson, Tobias Saule, Maksym Ivanov, Bruno~E. Schmidt, Zhanna Rodnova,
  George Gibson, Nora Berrah, and Carlos Trallero.
\newblock High-power femtosecond molecular broadening and the effects of
  ro-vibrational coupling.
\newblock {\em Optica}, 12(1):5--10, January 2025.

\end{thebibliography}

\end{document}